\documentclass[doublecol,figures]{epl2}

\usepackage{graphicx}
\usepackage{amssymb}

\renewcommand{\revision}[1]{{#1}}

\title{A simple self-organized swimmer driven by molecular motors}

\author{S. G\"unther\inst{1,2} \and K. Kruse\inst{1,2}}

\institute{
\inst{1} Saarland University, Department of Physics, Campus E2~6, 66123 Saarbr\"ucken, Germany\\
\inst{2} Max Planck Institute for the Physics of Complex Systems, N\"othnitzer
Str.~38, 01187 Dresden, Germany
}


\abstract{We investigate a self-organized swimmer at low Reynolds numbers. The microscopic 
swimmer is composed of three spheres that are connected by two identical active linker arms. 
Each linker arm contains molecular motors and elastic elements and can oscillate spontaneously. 
We find that such a system immersed in a viscous fluid can self-organize into a state of
directed swimming. The swimmer provides a simple system to study important aspects of the 
swimming of \revision{micro-organisms}.}

\begin{document}

\maketitle

\section{Introduction}

Micro-organisms often use flagella or cilia to move autonomously in an aqueous
environment~\cite{bray01}. In eukaryotes these hair-like appendages are internally
driven by molecular motors. The
motors set the appendage into oscillatory motion
leading to two- or three-dimensional beating patterns, which
propel the organism. Due to the complicated structure of cilia and
flagella, our understanding of swimming through these internally driven filaments 
is still far from being complete. This holds even more for effects caused by
hydrodynamic interactions between swimming \revision{micro-organisms}, which can lead to 
turbulent motion in bacterial suspensions~\cite{domb04} or to the formation of 
vortex arrays for ensembles of spermatozoa~\cite{ried05}. In the latter case,
the flagella propelling the spermatozoa have been observed to beat synchronously.
Synchronous beating can lead to metachronal waves~\cite{sleigh74} which are of 
biological importance for the swimming of the single-celled freshwater 
ciliate paramecium or the transport of mucus in our lungs.

\begin{figure}
\centering{\includegraphics[width=0.4\textwidth]{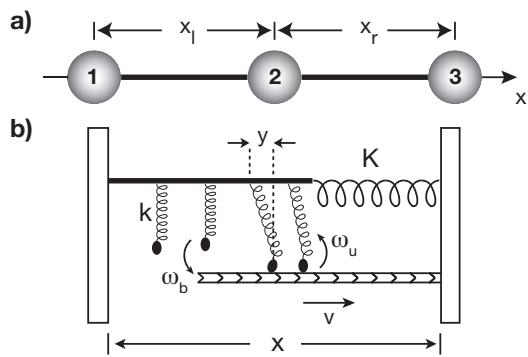} }
\caption{\label{fig:schema} \textbf{a)} Schematic representation of the three-sphere
system. \textbf{b)} Schematic representation of the dynamic linker element. It 
consists of a filament, a motor filament, and a linear spring with elastic modulus $K$. 
Motors bind to the filament with rate $\omega_{\mathrm{b}}$ and unbind with rate 
$\omega_{\mathrm{u}}$. Motors
bound to the filament move with velocity $v$ so as to shorten the linker length $x$.
Motors are connected to the backbone by springs of stiffness $k$ and extension $y$
along the $x$-axis. }
\end{figure}

Physical studies of the swimming of micro-organisms
have focused on the motion resulting from a given sequence of shape changes.
As the Reynolds numbers associated with the flow generated by \revision{such} 
swimmers are low, inertial effects do not play a role. Consequently, swimming requires
shape changes of more than one degree of freedom~\cite{purc77}. Different
swimmer geometries have been considered in this context, see, for example, 
Refs.~\cite{tayl51,purc77,shap87,purc97,naja04,avro05} the simplest being probably
the swimmer introduced by Najafi and Golestanian~\cite{naja04,gole08,golestanian08b,golestanian08a}, which consists
of three lined-up spheres with periodically varying distances, see Fig.~\ref{fig:schema}a.

The mechanisms underlying these shape changes, however, have received less attention
in the physical community.
Exceptions are provided by works on the oscillatory beating of 
flagella~\cite{cama99,cama00,riedel07}. There, the periodic motion is assumed to be 
generated by spontaneous mechanical oscillations of molecular motors coupled 
to elastic elements~\cite{brok75,juli97}. Combining the action
of motor molecules and the bending elasticity of a flagellum, wave patterns similar
to those observed experimentally are generated~\cite{cama99,cama00,riedel07}. 

In this work, we introduce a simple dynamical system consisting of three spheres that \revision{are arranged in
a line and that are connected by linkers. The linkers consist of molecular motors and elastic elements and their 
length can oscillate spontaneously.}
\revision{While }the swimmer is symmetric with respect to space-inversion \revision{the system can dynamically
break this symmetry and spontaneously swim directionally. We investigate the system's bifurcation diagram
and provide evidence that the transition to directional swimming shares striking similarities to transitions occuring in integrate-and-fire neurons.
} 

\section{The active three-sphere swimmer}
Consider three spheres, each of radius $R$ moving in a fluid of viscosity $\eta$ along the $x$-axis,
with neighboring spheres being linked by elements as illustrated in Fig.~\ref{fig:schema}b.
Their structure resembles that of sarcomeres, which are the
elementary force generating units of skeletal muscle, \revision{and consist of motors moving along a polar
filament combined with a linear spring. Such structures are known to be able to oscillate spontaneously~\cite{juli97},
a phenomenon that has been observed experimentally for muscle sarcomeres~\cite{okam88}.
Before investigating the dynamics of the three actively linked spheres, let us first describe the three-sphere
swimmer with externally imposed changed of the spheres' mutual distances.}

\begin{figure}
\centering{\includegraphics[width=0.4\textwidth]{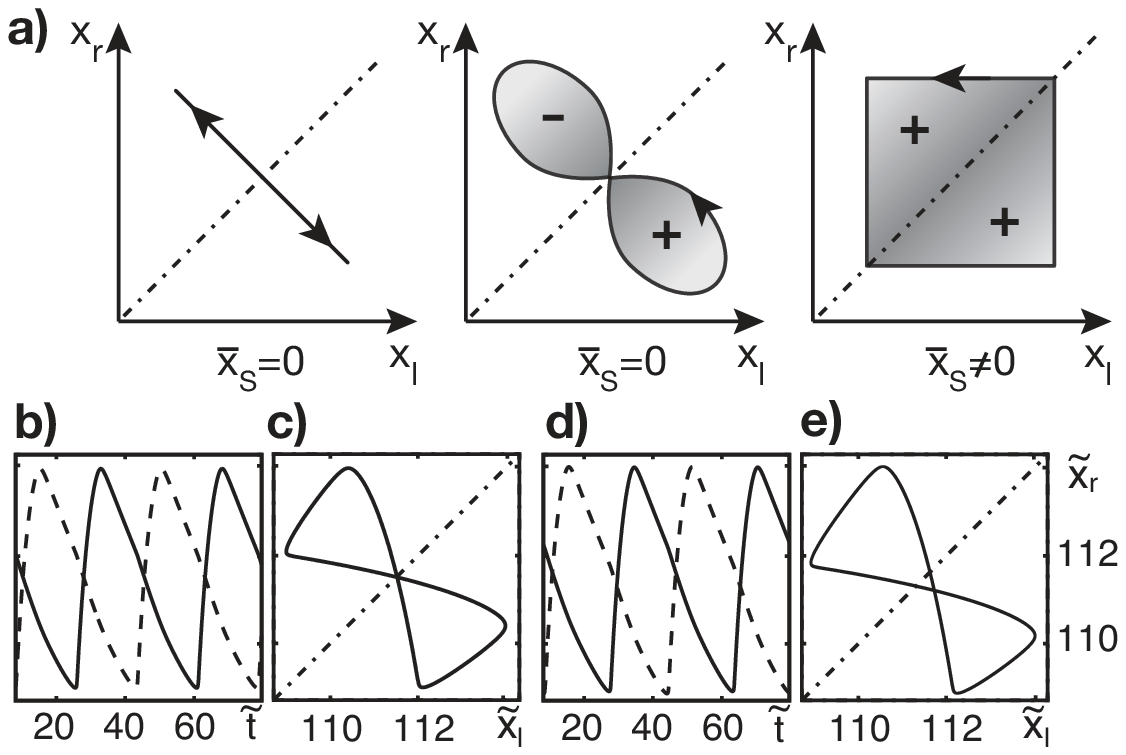}}
\caption{\label{fig:orbits} \textbf{a)} Schematic representation of possible orbits
in the $(x_{\:\!\mathrm{l}},x_{\mathrm{r}})$-plane of periodic motions of the three-sphere system.
Left: reciprocal motion, middle: non-reciprocal motion with $\bar x_{\mathrm{s}}=0$, right:
non-reciprocal motion with $\bar x_{\mathrm{s}}\neq0$. \textbf{b), c)} Left and right linker length 
as a function of time in the nonswimming state $\mathbf{O}_1$, see Fig.~\ref{fig:bifscene}, with 
$\xi\gtrsim\xi_{\mathrm{s}}$ and \textbf{d), e)} in the swimming state $\mathbf{Sw}$ with 
$\xi\lesssim\xi_{\mathrm{s}}$. Other parameters as in Fig.~\ref{fig:vvoneta}.}
\end{figure}

The three spheres are located at positions $x_i$, $i=1,2,3$ \revision{along the $x$-axis}.
At low Reynolds numbers, the forces 
$f_j$ acting on sphere $j$ with $j=1,2,3$ result in velocities $\dot{x}_i$ according 
to~\cite{batc67}
\begin{equation}
\label{eq:oseen}
\dot{x}_{i}=\sum_{j}H_{ij}\,{f}_{j}\quad.
\end{equation}
Here, the mobility tensor $H$ depends on the fluid's viscosity, the size of the
spheres, as well as the distances between them, and satisfies $H_{ij}=H_{ji}$. Using 
Eq.~(\ref{eq:oseen}) as well as the global force balance equation
$f_{1}+f_{2}+f_{3}=0$, the velocity of the center of mass of the system,
\revision{$\mathrm{d}x_{\mathrm{s}}/\mathrm{d}t\equiv\dot x_{\mathrm{s}}=(\dot x_{1}+\dot x_{2}+\dot x_{3})/3$}, can be expressed in terms
of the distances between the middle and the outer spheres, $x_{\:\!\mathrm{l}}$ and $x_{\mathrm{r}}$,
respectively. For periodically varying distances, one finds for the
displacement $\bar x_{\mathrm{s}}$ after one period $T$, 
\begin{equation}
\label{eq:xs}
\bar{x}_{\mathrm{s}}=\int_{0}^{T}\dot x_{\mathrm{s}}\; {\mathrm{d}}t=\int_{O}C(x_{\:\!\mathrm{l}},x_{\mathrm{r}}) \: {\mathrm{d}}x_{\:\!\mathrm{l}}\wedge
{\mathrm{d}}x_{\mathrm{r}}\quad,
\end{equation}
where $O$ denotes the oriented surface encircled in the $(x_{\:\!\mathrm{l}},x_{\mathrm{r}})$-plane
during one period of the motion, see Fig.~\ref{fig:orbits}a \revision{and also Refs.~\cite{purc77,shap87}}.
In this expression, $C$ is a field with
$C(x_{\:\!\mathrm{l}},x_{\mathrm{r}})=C(x_{\mathrm{r}},x_{\:\!\mathrm{l}})$ that only depends on the mobility 
tensor $H$. Thus the displacement is independent of the oscillation period. 
For the calculation of $C$, it is convenient to employ the Oseen approximation, which
consists in expanding $H$ in terms of 
$R/x_{\:\!\mathrm{l}}$ and $R/x_{\mathrm{r}}$ and retaining leading order terms only. In lowest order,
this corresponds to neglecting hydrodynamic
interactions between different spheres. In this case, $C(x_{\:\!\mathrm{l}},x_{\mathrm{r}})=0$, implying that
hydrodynamic interactions are essential for swimming of the three-sphere system.

If motion is reciprocal, that is, if the corresponding curve in the $(x_{\:\!\mathrm{l}},x_{\mathrm{r}})$-plane retraces
itself, then $\bar{x}_{\mathrm{s}}=0$ illustrating Purcell's Scallop theorem, see Fig.~\ref{fig:orbits}a left. 
Note, that a trajectory which is invariant under the exchange 
of left and right, i.e., space inversion, is necessarily reciprocal.
Examples of other periodic motions with vanishing net
displacement are given by $x_{\:\!\mathrm{l}}(t)=x_{\mathrm{r}}(t+T/2)$, see Fig.~\ref{fig:orbits}a middle. 
These solutions are invariant under space inversion and a simultaneous 
shift in time by $T/2$. A
simple form of $x_{\:\!\mathrm{l}}$ and $x_{\mathrm{r}}$ with $\bar{x}_{\mathrm{s}}\neq0$ was proposed by Najafi
and Golestanian~\cite{naja04}, see Fig.~\ref{fig:orbits}a right.

We now describe the dynamics of the linker elements depicted in Fig.~\ref{fig:schema}b, consisting of motors and elastic components. Similar systems have been studied in the contexts of spontaneous spindle~\cite{grill05} and muscle oscillations~\cite{gunt07}.

\revision{A linker element consists of motors moving directionally along a polar filament 
and a linear spring of stiffness $K$. 
The motors are attached to a common backbone by springs of stiffness $k$ and extension $y$.
In a sarcomere, the motors would be ensembles of myosin II that move on an actin filament, while
elastic elements are provided by various structural elements, e.g., the protein titin.
Motors bound to the filament move such that the linker shortens. An
unbound motor binds to the filament at rate $\omega_{\mathrm{b}}$.
Attached motors unbind from the filament at rate $\omega_{\mathrm{u}}$. These rates 
generally depend on the force applied to the motor. We assume that the force 
dependence is restricted to the unbinding rate. Motivated by Kramers' rate theory
we write $\omega_{\mathrm{u}}=\omega_{\mathrm{u}}^{0}\,\exp\left\{|f|\,a/k_{\:\!\mathrm{B}}T\right\}$, 
where $f=k\,y$ and $a$ is a microscopic length scale. Also the velocity of bound motors depends
on the applied force. For simplicity, we assume a linear force-velocity relation 
$v(f) = v_{0}\,(1-f/f_{0})$. Here, $f_{0}$ is the stall force and $v_{0}$ the
velocity of an unloaded motor.}

\revision{The dynamic equation for the length $x$ of the linker element connecting two spheres, say 1 and 2, 
$x=x_2-x_1$, is given by Eq.~(\ref{eq:oseen}), where the force $f_j$ on sphere $j$ is the sum of the elastic
and the motor forces generated by the linker. The elastic force $f_\mathrm{e}$ is given by $f_{\mathrm{e}}=-K(x-L_{0})$ with 
$L_0$ being the spring's rest length. To calculate the force exerted by the motors, $f_\mathrm{m}$, consider first a single
motor bound to the polar filament. If $y_n$ denotes the extension of the spring linking this motor to 
the backbone then the force it exerts is $-ky_n$.  The value of $y_n$ changes according to
\begin{equation}
\label{eq:ydot}
\dot y_n=v_n+\dot x\quad,
\end{equation}
where $v_n$ is the velocity of this motor on the filament. The total motor force then is $f_{\mathrm{m}}=-k\sum_{n=1}^{N}\sigma_{n}\,y_{n}$.
Here, $N$ is the overall number of motors  and $\sigma_{n}=1$ if
motor $n$ is bound and $0$ otherwise.We employ a mean-field
approximation, which consists in assuming that all bound motors have the same spring
extension, $y_n=y$ for all $n$. We determine the mean position by the product of the average motor speed 
multiplied with the average time they stay attached, $y=\dot y/\omega_{\mathrm{u}}$. For unbound motors, 
we assume fast relaxation, such that their 
distribution equals the equilibrium distribution implying $y=0$ for these motors. Under these conditions,
the fraction $Q$ of bound motors evolves according to~\cite{grill05}
}
\begin{equation}
\label{eq:qdot}
\dot Q=\omega_{\mathrm{b}}-(\omega_{\mathrm{b}}+\omega_{\mathrm{u}})\,Q\quad,
\end{equation}
while the total motor force is
\begin{equation}
\label{eq:motforcemf}
 f_{\mathrm{m}} = - N(x)\,Q\,k\,y\quad.
\end{equation}
Here, $N(x)=(\ell_{\mathrm{f}}+\ell_{\mathrm{m}}-x)/\Delta$ is the number of motors in the overlap region of
the filament and the motor backbone of lengths $\ell_{\mathrm{f}}$ and $\ell_{\mathrm{m}}$, respectively,
and $\Delta$ is the distance on the backbone
between adjacent motors.

\revision{Before considering the active three-sphere swimmer, let us present the dynamics of two actively 
linked spheres.}
The dynamic equations for this system are given by Eqs.~(\ref{eq:oseen}),
(\ref{eq:ydot})-(\ref{eq:motforcemf}), where the forces on the spheres 1 and 2 are
$f_{1}=-f_{2}=-f_{\mathrm{e}}-f_{\mathrm{m}}$. 
To lowest order the Oseen-expansion of $H$ gives
$H_{11}=H_{22}=(6\,\pi\,\eta\,R)^{-1}$, and $H_{12}=0$.
$H_{ii}$ is the Stokes friction of an isolated sphere.
\revision{In dimensionless form, the equations of motion read}
\revision{\begin{eqnarray}
\label{eq:eomhsx}
\xi\,\dot{\tilde{x}} &=& -2 \cdot \left[ N(\tilde{x})\,Q\,\kappa\,\tilde{y} + \tilde{x}-L \right] 
\equiv -2\,F(\tilde{x},Q,\tilde{y}) \: \; \\
\label{eq:eomhsq}
\dot{Q} &=& 1 - \left[1+\omega(\tilde{y})\right] \cdot Q  \\
\label{eq:eomhsy}
\dot{\tilde{x}} &=& \tilde{y} \cdot \left[ \omega(\tilde{y}) + \gamma \right] - 1 \quad,
\end{eqnarray}}
\revision{where the dimensionless parameters are given by $\xi=6\pi\eta R\,\omega_{\mathrm{b}}/K$, $L=L_{0}\omega_{\mathrm{b}}/v_{0}$, $\kappa=k/K$, 
$\omega=\omega_{\mathrm{u}}/\omega_{\mathrm{b}}$ and 
$\gamma=k v_{0} / (f_{0} \omega_{\mathrm{b}})$. Furthermore, $\tilde x=x\omega_\mathrm{b}/v_0$ and 
$\tilde y=y\omega_\mathrm{b}/v_0$, while time has been rescaled by $\omega_\mathrm{b}$}.

The equations have a stationary state $(\tilde x_{0},Q_{0},\tilde y_{0})$. For sufficiently short times,
a small perturbation $\delta \tilde x, \delta Q, \delta \tilde y$ will evolve as $\delta \tilde x(t)\propto\exp(st)$ 
and analogously for $\delta Q$ and $\delta\tilde y$.
It grows if $\Re(s)>0$. This condition can be expressed in terms of the system parameters\footnote{\revision{The instability occurs at
$\xi (1+\omega(\tilde{y})) + 1-Q\kappa\tilde{y}\tilde{\Delta} + [N(\tilde{x})Q\kappa(1+\omega(\tilde{y})-\tilde{y}\omega'(\tilde{y})]/[\omega(\tilde{y})+\tilde{y}\omega'(\tilde{y})+\gamma]=0 $, with $\tilde{\Delta}=v_0/\Delta \omega_{\mathrm{b}}$. Here, $\tilde{x}$, $Q$ and $\tilde{y}$ are taken
in the stationary state.} }
and implies $y_{0}\,\omega_{\mathrm{u}}'(y_{0})>\omega_{\mathrm{b}}+
\omega_{\mathrm{u}}(y_{0})$. The force dependence of the binding kinetics is thus essential
for an instability of the stationary state. The system oscillates if $\Im(s)\neq0$ at the instability. This
implies $Q_{0}\,k\,{y}_{0}<K\,\Delta$, that is, the increase of
the motor force by adding one motor in the overlap region needs to be smaller than the corresponding
increase of the restoring force by the elastic element to produce an oscillatory instability.

The oscillation mechanism can be understood intuitively as follows: As the motors shorten the
linker, the elastic restoring forces increases and, therefore, the force on the motors. This in turn
increases the unbinding rate of motors. When a few motors detach from the filament,
the remaining motors experience an even higher force, such that an avalanche of
motor unbinding events occurs. The elastic element then stretches the linker and 
the motors rebind, repeating the cycle. 

This behaviour shares similarities with the dynamics of integrate-and-fire model neurons~\cite{lapicque07}. 
There, the electric membrane potential of a neuron increases monotonically as a consequence of an input 
current. Upon reaching a threshold value, the potential is instantaneously reset and the loading process 
restarts. In the case of the linker element, the mechanical analog of neuron's electric potential is the internal 
stress of the linker. Loading occurs through the action of motors and resetting \revision{is accomplished by the 
spring $K$ after a} detachment avalanche of the motors.

Let us now return to the three-sphere system, see Fig.~\ref{fig:schema}, where the spheres are
connected by dynamic linker elements. The forces acting on the three spheres are
$f_{1}  =  -f_{\mathrm{e}}^{\mathrm{l}} -f_{\mathrm{m}}^{\mathrm{l}}$,
$f_{2}  =  -f_{1} - f_{3}$, and
$f_{3}  =  f_{\mathrm{e}}^{\mathrm{r}} +f_{\mathrm{m}}^{\mathrm{r}}$,
where the superscripts $l$ and $r$ distinguish between the left and the right linker,
respectively. 

As we have seen above, hydrodynamic interactions between different spheres are
essential for swimming of a three-sphere system. Compared to the two-sphere case, 
we thus now expand the mobility tensor to the next higher order.
This leads to $H_{ii}=(6\,\pi\,\eta\,R)^{-1}$, $i=1,2,3$,
$H_{12}=(4\,\pi\,\eta\,R)^{-1}R/x_{\:\!\mathrm{l}}$, $H_{23}=(4\,\pi\,\eta\,R)^{-1}R/x_{\mathrm{r}}$, and
$H_{13}=(4\,\pi\,\eta\,R)^{-1}R/(x_{\:\!\mathrm{l}}+x_{\mathrm{r}})$. \revision{The equations of motion are then 
given by Eqs.~(\ref{eq:eomhsq}) and (\ref{eq:eomhsy}) respectively for the left and right part of the swimmer. 
Equation~(\ref{eq:eomhsx}) is replaced by}
\revision{\begin{eqnarray}
\label{eq:eomsx}
\xi\,\dot{\tilde{x}}_{\:\!\mathrm{l},\mathrm{r}} &=& \left\{ \frac{3r}{\tilde{x}_{\:\!\mathrm{l},\mathrm{r}}}-2 \right\}
 F(\tilde{x}_{\:\!\mathrm{l},\mathrm{r}},Q_{\mathrm{l},\mathrm{r}},\tilde{y}_{\:\!\mathrm{l},\mathrm{r}}) 
\qquad \quad  \\ 
&&
+ \left\{ \frac{3r}{2} \left( \frac{1}{\tilde{x}_{\:\!\mathrm{l}}+\tilde{x}_{\mathrm{r}}}-
\frac{1}{\tilde{x}_{\:\!\mathrm{l}}}-\frac{1}{\tilde{x}_{\mathrm{r}}} \right)+1 \right\}
 F(\tilde{x}_{\mathrm{r},\mathrm{l}},Q_{\mathrm{r},\mathrm{l}},\tilde{y}_{\mathrm{r},\mathrm{l}})\nonumber
\end{eqnarray}
with $r=R\omega_{\mathrm{b}}/v_{0}$.} 
As before, the dynamic equations are invariant under 
space inversion, i.e., exchange of left and right.

\begin{figure}
\centering{\includegraphics[width=0.45\textwidth]{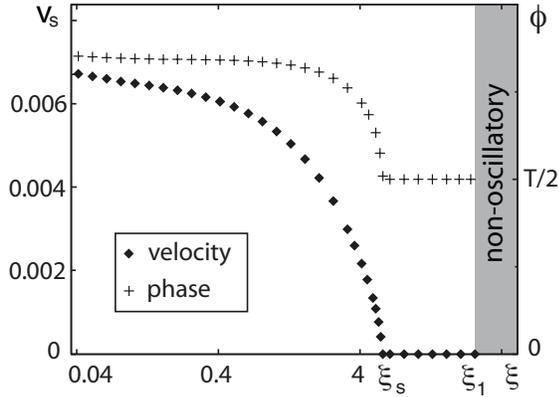}}
\caption{\label{fig:vvoneta}State diagram of the dynamic three-sphere swimmer. For $\xi>\xi_1$
the system is stationary. \revision{Filled diamonds indicate the mean velocity of the swimmer's 
center of mass.} For $\xi_\mathrm{s}<\xi<\xi_1$ it oscillates, but does not move on 
average, $v_\mathrm{s}=0$. 
For $\xi<\xi_\mathrm{s}$, $v_\mathrm{s}>0$. 
Crosses indicate the relative phase between the oscillations of the two
linker elements. 
Equations~(\ref{eq:eomhsq})-(\ref{eq:eomsx})
were solved numerically with $L=137.5$, $\kappa=400$, $\omega_{0}=0.5$, $\gamma=9.1$, $\tilde{\Delta}=0.16$ and $r=11$.}
\end{figure}

\revision{
Figure~\ref{fig:vvoneta} summarizes the three-sphere system's behavior with the active linker elements as the parameter $\xi$ is changed. This can be achieved, for example, by changing the viscosity of the surrounding
fluid or the binding rate of the motors.  
Having $\xi>\xi_1$, the system settles into a stationary state. For $\xi<\xi_1$, the linkers spontaneously oscillate 
in length. The oscillations of the two linker elements are identical but shifted with respect to each other by half 
a period. As is illustrated in Fig.~\ref{fig:orbits}b,c, this symmetry implies that, for low Reynolds numbers, the 
swimmer's center 
of mass ${x}_{\mathrm{s}}$ remains stationary on average. There is a second critical value $\xi_\mathrm{s}$, 
at which the relative phase shift between the two oscillating linkers $\Phi$ starts to deviate from half a period, such that
the system spontaneously swims if $\xi<\xi_\mathrm{s}$. In contrast, the form of the oscillatory motion 
of each of the two linker elements essentially does not change. By dynamically breaking the space-inversion symmetry, the corresponding 
orbit in the $(x_{\:\!\mathrm{l}},x_{\mathrm{r}})$-plane is not symmetric with respect to the line 
$x_{\:\!\mathrm{l}}=x_{\mathrm{r}}$,
see Fig.~\ref{fig:orbits}d,e, hydrodynamic interactions allow in average for
directed motion of the three-sphere system.

Decreasing the value of $\xi$ further speeds up the swimmer. The increase in the swimming velocity is the consequence of two effects. On one hand, the relative phase shift 
gets larger and, on the other hand, the amplitude of the oscillation increases, both leading 
to larger areas encircled by the orbit in the $(x_{\:\!\mathrm{l}},x_{\mathrm{r}})$-plane. For 
still lower viscosities one expects eventually a decrease in the velocity such that swimming stops
at $\xi=0$. The dynamic equations presented here do not capture this effect as
their applicability is restricted to the regime of low Reynolds numbers. 

\begin{figure}
\centering{\includegraphics[width=0.4\textwidth]{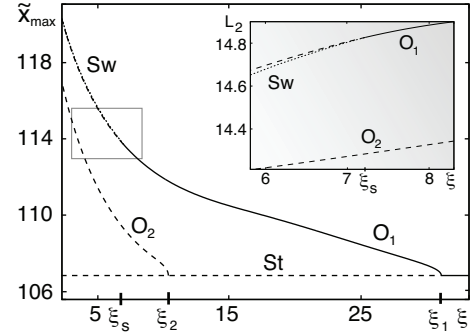}}
\caption{\label{fig:bifscene} Bifurcation scenario represented by the linker amplitude $\tilde{x}_\mathrm{max}$ for varying parameter $\xi$. Solid and dotted lines indicate stable, dashed lines unstable states. For $\xi<\xi_1$ the stationary state $\mathbf{St}$ becomes unstable via a Hopf-bifurcation. The emerging oscillatory state $\mathbf{O}_1$ loses stability at $\xi=\xi_{\mathrm{s}}$ and the swimming solution $\mathbf{Sw}$ (dotted line) appears. Furthermore, at $\xi=\xi_2$, a second oscillatory Hopf-mode $\mathbf{O}_2$ appears. Inset: Representation of the swimming bifurcation in terms of the $L_2$-norm (see text).}
\end{figure}

By the symmetry of the swimmer, with each moving state coexists a state 
moving into the opposite direction. The actual direction of motion is determined by the initial conditions.
Let us also note that changing the parameters $\gamma$ or $r$ can also lead to transitions from 
the stationary to the non-swimming oscillatory state or from the non-swimming oscillatory to the swimming 
state. We did not find two subsequent bifurcations leading from the stationary to the swimming state, though.
}

We will now take a closer look at the swimming transition. The bifurcation diagram of the swimmer 
\revision{in terms of the oscillation amplitude $\tilde{x}_\mathrm{max}$} is presented in Fig.~\ref{fig:bifscene}. The stationary 
state $\mathbf{St}$ loses stability through a 
supercritical Hopf-bifurcation at \revision{$\xi=\xi_1$}. The newly generated oscillatory state $\mathbf{O}_1$ 
satisfies $\tilde x_{\mathrm{l}}(\tilde t)=\tilde x_{\mathrm{r}}(\tilde t+T/2)$, where $T$ is the dimensionless
oscillation period. At \revision{$\xi=\xi_{\mathrm{s}}$}, $\mathbf{O}_1$ loses 
stability through a pitchfork bifurcation, see Fig.~\ref{fig:vvoneta}, resulting in the swimming state 
$\mathbf{Sw}$. Here, $\tilde x_{\mathrm{l}}(\tilde t)=\tilde x_{\mathrm{r}}(\tilde t+\phi)$, where the 
relative phase $\phi \neq T/2$. For \revision{$\xi<\xi_{\mathrm{s}}$}, the states $\mathbf{O}_1$ and $\mathbf{Sw}$ 
have almost the same amplitude. 
Therefore, we display in the inset of Fig.~\ref{fig:bifscene} a representation of the bifurcation in terms 
of the $L_2$-norm, $T^{-1}\int_0^T \mathrm{d}\tilde t \left( \tilde x_{\mathrm{l}}^2+\tilde x_{\mathrm{r}}^2+
Q_{\mathrm{l}}^2+Q_{\mathrm{r}}^2+\tilde y_{\mathrm{l}}^2+\tilde y_{\mathrm{r}}^2 \right)$.
For a similar value of \revision{$\xi$} a second non-swimming state $\mathbf{O}_2$ bifurcates from the stationary 
state at \revision{$\xi=\xi_2$}. This state, which is unstable for all values of \revision{$\xi$}, satisfies $\tilde x_{\mathrm{l}}(\tilde t)=
\tilde x_{\mathrm{r}}(\tilde t)$. 

The appearance of the state $\mathbf{O}_2$ together with an inspection of the corresponding isotropy lattice might suggest mode interaction to be at the origin of the symmetry transition. However, the frequencies of $\mathbf{O}_1$ and $\mathbf{O}_2$ are different at \revision{$\xi_{\mathrm{s}}$}, which argues strongly against mode interaction.

While we are currently lacking a thorough understanding of the swimming transition, we would like to 
point out that this transition is reminiscent of a phenomenon occuring for two coupled integrate-and-fire 
neurons~\cite{vreeswijk94}. \revision{The membrane potential of an integrate-and-fire neuron is increased
at a constant rate and resetted when a certain threshold voltage is reached. Upon resetting, the neuron sends
an electric pulse to other neurons it is connected to and influences their membrane potentials. }
For broad pulses, the two coupled neurons \revision{settle into a state in which they }fire periodically with a 
relative phase shift of \revision{half the period. }
For short enough pulses, however, this state is unstable and a new value of the phase-shift is 
generated~\cite{vreeswijk94}. For the swimmer, in analogy to the electrical pulses for the neurons, 
mechanical pulses are transmitted netween the two linkers. A mechanical pulse is released by one linker 
when it relaxes as a consequence of a detachment avalanche of the motors. In that case, the duration of 
the pulse changes, because of alterations of the duration of the linkers expansion phase that effectively 
transmitts the mechanical pulse. 
\revision{Decreasing $\xi$ leads to faster stretching of the linkers. In terms of physical parameters, 
decreasing $\xi$ either corresponds to reducing friction forces on the beads, to decreasing the binding 
rate of motors, 
or to strengthening the elastic elements. As for the two-neuron system, a sharper pulse can be sufficient 
to cause dephasing~\cite{timme08}, which implies swimming.}

\revision{
The equations of motion (\ref{eq:eomhsq})-(\ref{eq:eomsx}) are of the form of a 
singularly perturbed system with $\xi$ as the small parameter. Such systems are known to present 
strong deformations of a 
limit cycle close to a Hopf-bifurcation, where the deformation occurs for exponentially small 
parameter changes~\cite{benoit81}. In certain parameter regions, we indeed find that 
the Hopf-bifurcation is followed by such a deformation which is called a Canard explosion. 
Accompanying with the Canard phenomenon, we find also excitable dynamics for the swimmer.}

~\revision{In addition to the Canard explosion, the singularly perturbed system (\ref{eq:eomhsq})-(\ref{eq:eomsx})
shows  dynamic behavior that is reminiscent of the dynamics 
associated with global bifurcations. In this case, the spontaneous
oscillations of the linker are still generated by a Hopf bifurcation. The dynamics close to the bifurcation, however,
shows characteristics of the dynamics close to a saddle-node infinite period bifurcation 
(SNIPER). A SNIPER bifurcation is characterized by two fixed points, a saddle and a node, that collide, which 
leads to oscillations around a third, unaffected fixed point, namely, an unstable node~\cite{strogatz94}. 
In the present case, this limit cycle is symmetric with respect to space-inversion. When decreasing the control parameter $\xi$
further, the limit cycle splits up into two limit cycles that are mirror images of each other upon space inversion.
This transition is similar to an anti-gluing bifurcation~\cite{coullet84}. A detailed analysis of these bifurcations
will be given elsewhere.}

\section{Discussion}
In this work, we have presented a simple self-organized swimmer. It consists of three aligned spheres 
connected by active linker elements that can oscillate spontaneously. \revision{By decreasing the 
dimensionless parameter $\xi$ below a critical value, which can be achieved either by reducing the
the viscosity of the surrounding liquid or the motors' binding rate or by increasing the stiffness $K$ of the 
elastic element, the system oscillates spontaneously.}
While this state is not associated with an average displacement, a subsequent bifurcation encountered 
when decreasing $\xi$ further, leads to directed swimming. 
\revision{Changing other system parameters can either lead to spontaneous oscillations or to swimming 
but has not been found to induce both transitions subsequently.}

Could the self-organized three-sphere be realized experimentally? One approach could be based on muscle 
sarcomeres as linker elements. As mentioned above, sarcomeres can oscillate spontaneously~\cite{okam88}. 
Using parameter values appropriate for sarcomeres\footnote{\revision{$\ell_{\mathrm{f}}=0.6\,\mu$m, $\ell_{\mathrm{m}}=0.75\,\mu$m, $\Delta=56\,$nm, $k=4\,$pN/nm, $K=0.01\,$pN/nm, $L_{0}=1.25\,\mu$m, $v_{0}=2\,\mu$m/s, $f_{0}=4\,$pN, $k_{\mathrm{B}}T=4\,$pN$\,$nm, $a=3\,$nm, $\omega_{\mathrm{b}}=220$/s and $\omega_{\mathrm{u}}^{0}=120$/s.}} and assuming that they link spheres with a diameter of 
$R=0.1\,\mu$m, we find swimming velocities on the order of $v_{\mathrm{s}}\approx1\,\mu$m/min. This has to 
be compared to the diffusion constant. For a sphere with a diameter of $0.1\mu$m we find in water at room 
temperature $D\approx 2 \,\mu$m$^2$/s. On a distance comparable to the swimmer size $L\approx3\,\mu$m, 
diffusion thus dominates the motion $L v_{\mathrm{s}}/D=0.025$. In order to reach bigger swimming speeds, 
the oscillation frequency could be increased. To this end stiffer springs and stronger motors would be necessary.
\revision{Another way to increase the swimming speed would be through modifications of the swimmer design.
For example, it has been argued that the pushmepullyou swimmer~\cite{avro05} is faster than the three-sphere
swimmer.}

Our self-organized swimmer can be used to study the effects of hydrodynamic coupling between 
micro-swimmers. So far, studies of this subject have largely neglected the mechanism used by micro-swimmers 
to generate the shape changes necessary for swimming. Preliminary results suggest interesting
synchronization effects that greatly alter the system's behavior compared to swimmers with fixed 
sequences of shape changes. Obviously, it would also be interesting to extend our approach to study 
the dynamics of cilia and flagella.

\acknowledgments
We thank A. Hilfinger and E.~M. Nicola for useful discussions.

\end{document}